\documentclass[12pt,letterpaper]{article}
\usepackage[utf8]{inputenc}
\usepackage[english]{babel}
\usepackage{amsmath}
\usepackage{amsfonts}
\usepackage{amssymb}
\usepackage{graphicx} 
\usepackage{listings}
\usepackage{color}
\usepackage[left=2cm,right=2cm,top=2cm,bottom=2cm]{geometry}
\allowdisplaybreaks

\begin{document}

\large
\begin{center}
	\bf{The Role that Gaiters, Masks and Face Shields Can Play in\\ Limiting the Transmission of Respiratory Droplets.} 
\end{center}
\vspace{20pt}

\begin{center}
Gavin A. Buxton, Science Department, Robert Morris University.\\
  
Marcel C. Minutolo, Management Department, Robert Morris University.

\end{center}
\normalsize

\vspace{40pt}

\begin{center}Abstract\end{center}

The efficacy of face masks, neck gaiters and face shields are predicted and contrasted.
In particular, a Lattice Spring Model of a neck gaiter serves as the input to a Lattice Boltzmann simulation.
The Lattice Boltzmann method is used to capture the fluid dynamics both through and around various face coverings. 
The evaporation and transport of respiratory droplets is simulated in this flow field, and the efficacy of the different face coverings at stopping the spread of respiratory droplets are elucidated and contrasted.
In agreement with recent experimental studies, we predict that neck gaiters are at least as effective as cloth masks, and both are far superior to a face shield.

\newpage
\section{Introduction}

The U.S. Centers for Disease Control and Prevention (CDC) and the World Health Organization (WHO) are now recommending face masks be used by the general public and it is believed that, in conjunction with other mitigating behaviors, the wide-spread use of face masks can reduce community spread of Covid-19 \cite{cheng2020wearing}.  
Given limited personal protective equipment (PPE), and the needs of healthcare workers on the front-line of the battle against this pandemic, cloth face-masks have been widely recommended for the general public (e.g. \cite{howard2020face}). Mukerji \emph{et al.} use the term `mask' to include a class of standard surgical masks which are not specially designed to protect the wearer from aerosol transmission \cite{mukerji2015infectious}. Further, they refer to `respirators' as personal protective equipment that is designed to filter and fit such that it prevents the transmission of the aerosol droplets. Included in the category of respirators are the N95, high-efficiency particulate air, powered air purifying respirator, dust-mist, and the dust-mist-fume types of respirators. Given the need and demand for the respirators by healthcare workers and first responders and the limited supply thereof, public good suggests that those items ought to be prioritized for these front-line workers. Additionally, the cost difference between masks and respirators is such that it may be prohibitive for some to acquire the later. Prior to wearing the respirator, it is recommended that the user undergo fit testing prior to use \cite{chughtai2013availability}, whereas fit testing for masks is not necessary. Given the pressure put on the supply and the challenges that may exist for training how to fit respirators, the use of masks for the general public is more feasible.\\

With the shortage and expense of respirators, the widespread adoption of single use masks was an initial response of some within the general public. However, the disposal of single use masks posses a potential environmental catastrophe \cite{fadare2020covid, roberts2020coronavirus}. To this end, the use of cloth face-masks are more environmentally responsible than single-use face masks \cite{klemes2020energy, roberts2020coronavirus}, and have been shown to be effective in reducing the transmission of other respiratory diseases \cite{spooner1967history}. However, controversy remains concerning US public opinion on the efficacy of face-masks, neck gaiters and face shields for mitigating the spread of Covid-19 \cite{wong2020covid}. The controversy over the efficacy lay in the doubt over the proper use of the PPE, lack of understanding of the transmission, and / or beliefs concerning the ability of the PPE to stop the movement of the droplets.\\

Respiratory droplets are known to contain potential pathogens \cite{hare1964transmission, wells1934air, corbett2003growing}, and are a major contributor to annual influenza epidemics \cite{solomon2020influenza}. SARS-CoV-2 is also believed to be spread through respiratory droplets \cite{Lai2020, huang2020}. The droplets are emitted from the humid respiratory tract of an infected individual, travel outside of the body, before potentially landing in the respiratory tract of a susceptible individual.\\

Limiting this mechanism of spreading is particularly important for the current COVID-19 pandemic. In particular, the presence of asymptomatic individuals \cite{asadi2019aerosol} could result in approximately 40\% to 45\% of SARS-CoV-2 infections \cite{oran2020prevalence}, which is why the CDC recommend everyone wears a mask regardless of symptoms \cite{howard2020face}. One study has estimated that the economic benefit of each additional cloth mask worn ranges from \$3,000 to \$6,000 \cite{abaluck2020case} stemming,  in part, from the recovery of the estimated \$60,000 per capita mortality risk \cite{greenstone2020social}.\\

The movement of respiratory droplets from one individual to another, however, is a complex phenomena, affected by the evaporation, the dispersion, and the deposition of the droplets \cite{wells1934air}. The size of the droplets play a critical role. As the droplets of various sizes are emitted they immediately start to evaporate towards smaller droplet nuclei. This is important because larger particles are more likely to deposit on various surfaces, turning them into potentially infectious fomites, while smaller droplets might linger in the air and follow air current more closely. Therefore, airborne transmission is likely to depend on the background air flow \cite{lowen2006guinea, ai2018airborne} and ventilation \cite{giovanni2020transmission}. Protected in the droplet nuclei \cite{huang2020}, it is known that a variety of pathogens can survive for prolonged periods in the air \cite{harper1961airborne}, including SARS-CoV-2 \cite{van2020aerosol}.\\
	
In terms of the use of personal protective equipment, the role that airborne transmission plays in the spread of COVID-19 has been controversial as this might necessitate more stricter protocols, such as the use of airborne isolation infection rooms and fit-tested N95 face-piece respirators \cite{Leung2019}. The relative importance of airborne transmission versus larger droplet deposition is also expected to play a large role in the efficacy of different masks and face coverings.\\

Larger droplets are unlikely to travel far before gravity deposits them on nearby surfaces. This is one of the reasons that social distancing, especially when individuals are facing one another \cite{ai2018airborne}, is an important measure for decreasing the risk of transmission. However, SARS-CoV-2 has been found to persist for days on surfaces \cite{van2020aerosol} (more so on smooth surfaces than desiccating porous surfaces \cite{huang2020}) meaning that fomites are also a large contributing factor in the transmission of COVID-19. While frequent hand-washing will indubitably reduce the risk of transmission from a fomite to a susceptible individual \cite{jefferson2020physical}, the use of masks is also expected to reduce the formation of fomites by limiting the emission of large droplets from infected individuals \cite{huang2020}.\\

The location within the respiratory track from where the respiratory droplets originate will dictate the concentration of pathogens contained within the respiratory droplets \cite{wei2015enhanced}. In particular, different pathogenic organisms can be more concentrated at different locations along the respiratory tract. For instance, the instability and fluid film rupture of mucus layers along the respiratory bronchioles are thought to be responsible for smaller respiratory droplet formation \cite{johnson2009mechanism}. Large concentrations of influenza pathogens have been found in smaller droplets \cite{yan2018infectious, Lindsley2010}, and this maybe exacerbated with the lower respiratory tract infections associated with SARS-CoV-2 \cite{asadi2019aerosol}. Alternatively, larger respiratory droplets are more likely to be produced within the large airways of the upper respiratory tract and the oral cavity \cite{wei2015enhanced}. These droplets are emitted at velocities of between 6 and 22 m/s during coughing and between 1 and 5 m/s during regular breathing \cite{wei2015enhanced, chao2009characterization, kwon2012study, wei2016airborne}.\\

Significant inconsistencies have been reported in regards to the size distribution of respiratory droplets \cite{duguid1946size, chao2009characterization, gupta2010characterizing, zhang2015documentary, Liu2017, asadi2019aerosol}, but the radius of respiratory droplets is typically found to range from 1 $\mu$m to 100 $\mu$m, with a greater number of droplets at a radius of between 5 $\mu$m and 10 $\mu$m \cite{duguid1946size}. This is important, not only as to how it pertains to the efficacy of face masks, but because smaller respiratory droplets are more likely to penetrate further into the respiratory tract \cite{licht1972movement, stahlhofen1983deposition} where a smaller number of pathogens are believed to be required to cause infection \cite{Thomas2013}. For this reason advocating for the wide spread use of masks, that may stop the emission of respiratory droplets at the source, is considered an important mitigating strategy during this pandemic.\\

While some political and social resistance to wearing masks exists, it is widely agreed that wearing masks will reduce community spread without significant social or economic impacts \cite{feng2020rational}. A large number of materials have been considered for use in constructing homemade masks, and the filtration efficiency of these materials has exhibited considerable variability \cite{davies2013testing, bagherifiltration, lindsley2020efficacy, konda2020aerosol, teesing2020there, kahler2020fundamental, aydin2020performance, greenhalgh2020face}. Furthermore, neck gaiters (elastic fabric tube that one can wear around the face) have emerged as a popular face covering, with limited studies concerning their effectiveness \cite{lindsley2020efficacy}.  In particular, the role of masks and coverings at limiting the release of respiratory droplets \cite{wei2016airborne, dbouk2020respiratory}, or diverting smaller droplets and reducing their forward momentum \cite{huang2020}, is widely accepted \cite{leung2020respiratory}. However, the role of masks and coverings in protecting susceptible individuals from the respiratory droplets of others (already in the air) is more controversial \cite{davies2013testing, wei2016airborne, kahler2020fundamental}.\\

The filtration efficiency depends on the size of the droplets with large droplets, that might be more likely to deposit on nearby surfaces, more effectively blocked by fabrics \cite{drewnick2020aerosol, aydin2020performance, leung2020respiratory}. Therefore, masks are expected to be effective in reducing the formation of fomites. However, multiple layers of fabric have been found to substantially increase the filtration efficiency of mask materials \cite{drewnick2020aerosol, zangmeister2020filtration}, with the first layer reducing the velocity of the droplets and increasing the ability of the second layer to filter the droplets \cite{aydin2020performance}. In addition, triboelectrically charged fabrics have been found to significantly enhance filtration efficiencies \cite{zhao2020household}, and this is expected to favor artificial materials. Neck gaiters, commonly made from polyester fabric, are high in the triboelectric series \cite{zou2019quantifying}, and may benefit from triboelectrification. However, the reason neck gaiters are popular (especially among runners) is their breathability, and it is common for materials with high breathability to have reduced filtration efficiency \cite{aydin2020performance}. The reduction in velocity through a more impermeable mask would be expected to be larger \cite{tang2009schlieren}, but the velocity through any fabric would be expected to be significantly reduced \cite{verma2020visualizing, aydin2020performance}. Furthermore, the effect of leakage due to poor mask to face fitting will degrade the collection of respiratory droplets by the face mask, with an area of leakage of only 2\% the area of the mask causing a mask efficiency reduction of 66\% \cite{dbouk2020respiratory}. The air flow will preferentially follow the path of least resistance \cite{drewnick2020aerosol}, and smaller droplets will likely follow this path around the mask \cite{dbouk2020respiratory}, regardless of filtration efficiencies.\\

Recent computer simulations have been used to elucidate the role of fluid mechanics on the efficacy of masks. In particular, Dbouk and Drikakis \cite{dbouk2020respiratory} investigated the role of leakage on the efficiency of masks. They predicted that most large droplets would become trapped at the mask surface, while smaller droplets are more likely to follow the air flow. Kumar \emph{et al.} simulated a cloth mask as an isotropic porous medium with an imperfect fit, and captured the flow through and around a mask, along with the spatial spread of droplets \cite{kumar2020utility}. The effect of the mask was found to reduce velocities, redirect air flow around the mask, and retain the majority of the ejected droplets. Here, we contribute to these emerging studies by simulating both the air flow through and around a face mask, a neck gaiter and a face shield, and the spatial evolution of emitted respiratory droplets within these flow fields. The next section details the methodology used in this study. Following the methodology, we present our findings and then discuss the implications.

\section{Methodology}

To capture the ability of face masks, neck gaiters and face shields at containing respiratory droplets we combine a number of simulation techniques. First a Lattice Spring Model (LSM) os elastic mechanics is used to simulate neck gaiter configurations. Second the Lattice Boltzmann (LB) method is used to simulate the fluid mechanics through and around porous face coverings (and around an impermeable face shield). Lastly, the spatial evolution of evaporating respiratory droplets are simulated in the flow fields and the collection efficiency of the different face coverings contrasted.

\subsection{Lattice Spring Model}

The Lattice Spring Model (LSM) is a coarse-grained simulation of elasticity and fracture mechanics. 
In particular, the coarse-grained components of the model (elastic springs) mirror atomistic processes (interatomic attractions) such that the correct continuum behavior (linear elasticity) emerges.
The mechanics of the neck gaiter are captured using this coarse-grained spring
model (with spring constants $k_{i,j}$ connecting nodes $i$ and $j$) representation of a two-dimensional surface. The energy associated with the system is of the form
\begin{equation}
A = \sum \frac{k_{i,j}}{2} (r_{i,j} - r^o_{i,j})
\end{equation}
where $r_{i,j}$ is the distance between nodes $i$ and $j$, and $r^o_{i,j}$ is the equilibrium distance. 
A square lattice is implemented with nodes connected to their nearest and next-nearest neighbors.
While more complicated Hamiltonians can be implemented, this Hookean Spring Model can capture the deformation of a fabric mask with an elastic modulus of $8 k/3$ and a Poisson's ratio of $1/3$ in the square lattice considered here.
No bending motions are considered, and the fabric is free to fold or ``bunch up'' without any energy penalty.
The equilibrium position of the LSM nodes is found by minimizing the energy of the system.\\

The facial dimensions were obtained from the National Institute for Occupational Safety and Health (NIOSH) anthropometric survey \cite{zhuang2010digital} in the form of a series of polygons that define averaged facial characteristics.
The displacement of the nodes are confined at the boundaries (either side of the head) and the nodes are updated iteratively to reduce the elastic energy, whilst ensuring the nodes do not pass through the polygons that define the face.
In this manner, the Lattice Spring Model captures the deformation of a neck gaiter as it stretches around the face of a wearer.

\subsection{Lattice Boltzmann Method}

The Lattice Boltzmann (LB) method is another coarse-grained simulation technique whose emergent behavior comes from coarse-grained components that mirror microscopic phenomena \cite{chen1998lattice}.
In particular, the Lattice Boltzmann method generally consists of two steps. First, a streaming step representing the advection of the particles in the fluid.
\begin{equation}
f^*_i(\mathbf{x} + \mathbf{c}_i \Delta t, t) = f_i(\mathbf{x}, t)
\end{equation}
Second, a collision step that captures the particle collisions that take place during the
movement of the particles in the fluid.
\begin{equation}
f^{**}_i(\mathbf{x}, t + \Delta t) = f^*_i(\mathbf{x}, t) - \frac{1}{\tau} \left(f^*_i(\mathbf{x}, t) - f_i^{eq}(\mathbf{x}, t) \right)
\end{equation}
The collision step relaxes the distribution functions towards an equilibrium distribution function of the form
\begin{equation}
f_i^{eq}(\mathbf{x}, t) = \rho w_i \left( 1 + 3 \frac{\mathbf{u} \cdot \mathbf{c}_i}{c} + \frac{9}{2}  \frac{\left(\mathbf{u} \cdot \mathbf{c}_i\right)^2}{c^2} - \frac{3 u^2}{2c^2}\right)
\end{equation}
The density is $\rho = \sum_i f_i$ and velocity is found from $\rho \mathbf{u} = \sum_i f_i \mathbf{c}_i$. 
The lattice weight factors for the D3Q19 model are $w_0 = 1/3$, $w_{1 \rightarrow 6} = 1/18$, and $w_{7 \rightarrow 18} = 1/36$. 
The lattice sound speed is $c = \Delta x/\Delta t$ and the viscosity of the fluid is $\mu = (2 \tau - 1) c^2 \Delta t / 6 \rho$.
Here $\Delta t = \Delta x = 1$, and the density is initially set to unity \cite{he1997theory, thurey2009stable, gaedtke2018application, buxton2005newtonian, buxton2006computational}.\\

To capture the effects of turbulence on the fluid flow the Smagorinsky subgrid
turbulence model is implemented, which essentially modifies the viscosity according to the Reynolds stress tensor
to account for subgrid scale vortices. This modifies the relaxation time in the relaxation process of the lattice Boltzmann equations such that each node of the lattice
relaxes at different rates \cite{si2015study}.
The modified relaxation time $\tau_s$
is computed as
\begin{equation}
\tau_s = 3 (\nu + C^2 S) + \frac{1}{2}
\end{equation}
where $C$ is Smagorinsky constant, for which a value of 0.04 is chosen, and $S$ is the local strain tensor.
\begin{equation}
S = \frac{1}{6C^2} \left( \sqrt{\nu^2 + 18C^2 \sqrt{\Pi_{\alpha,\beta}\Pi_{\alpha,\beta}}} - \nu\right)
\end{equation}
where the tensor $\Pi_{\alpha, \beta}$ is obtained for each cell from the second
moment of the non-equilibrium parts of the distribution functions.
\begin{equation}
\Pi_{\alpha, \beta} = \sum_{i=1}^{19} \mathbf{c}_{i\alpha} \mathbf{c}_{i\beta} (f_i - f_i^{eq})
\end{equation}
$\alpha$ and $\beta$
each run over the three spatial dimensions, while $i$ is the
index of the particle distribution functions for the D3Q19 model \cite{thurey2009stable}.\\

No-slip boundary conditions are implemented at the solid nodes, and partially at porous nodes that represent the neck gaiter and mask, with the following third step for the particle distribution function \cite{li2014lattice}.
\begin{equation}
f_i(\mathbf{x}, t+\Delta t) = f^{**}_i(\mathbf{x}, t+\Delta t) + n_s \left( f^{**}_{\acute{i}}(\mathbf{x}, t) - f^{**}_i(\mathbf{x}, t+\Delta t)\right)
\end{equation}
Where $n_s$ is the fraction of solid at a given node. When $n_s = 1$ the above recovers the bounce-back boundary conditions, where incoming fluid particles are simply reflected in
the opposite direction during the collision step; $\acute{i}$ is the
opposite direction of $i$. Von Neumann boundary conditions are implemented at the boundaries \cite{junk2008outflow}.  

\subsection{Droplet Evaporation and Transport}

In the current model we simulate individual droplets as they emerge from the mouth, shrink as the fluid evaporates, are subject to gravity and buffered by the surrounding air flow \cite{buxton2020spreadsheet}.  
The diameter of a droplet changes as the water evaporates up until a critical time, $t_{crit}$, when the droplet diameter has reached its minimum, $d_{min}$  \cite{Wallace, Shaman2009, Halloran2012}. 
\begin{equation}
d = \begin{cases} d_0 \sqrt{1 - \beta t} &\mbox{if } t \leq t_{crit} \\
d_{min} & \mbox{if } t > t_{crit}  \end{cases}
\end{equation}
where $t$ is time after the droplet is emitted, and $\beta$ is the evaporation rate given by
\vspace{10pt}
\begin{equation}
\beta = \frac{8 D (P_{sat} - P_{\infty})}{d^2 \rho R_v T}
\end{equation}
where $D$ is the molecular diffusivity of water vapor, $P_{sat}$ and $P_{\infty}$ are the saturation and ambient water vapor pressures, $R_v$ is the specific gas constant for water, and $T$ is temperature.\\

The minimum diameter is considered to be 44\% of the original diameter, or $d_{min}/d_0 = 0.44$ \cite{Nicas2005, Halloran2012}.
However, others have argued for smaller values, or that the minimum diameter might depend on the ambient humidity \cite{Liu2017}.  
$R_v = 461.52\,\text{J/(kg}\, \text{K})$ is the specific gas constant for water and the saturation water vapor pressure in Pascals can be obtained from the Buck equation \cite{Buck1981}.
\vspace{10pt}
\begin{equation}
P_{sat} = 611.21 \exp\left( \left(19.843 - \frac{T}{234.5}\right)\left(\frac{T - 273.15}{T - 16.01}\right)\right)
\end{equation}
The ambient water vapor pressure in Pascals is given by
\vspace{10pt}
\begin{equation}
P_{\infty} = P_{sat}\, \frac{RH}{100\%}
\end{equation}
where $RH$ is the relative humidity and the molecular diffusivity (in m$^2$/s) of water vapor in air is given by
\vspace{10pt}
\begin{equation}
D = 2.16\times10^{-5} \left(\frac{T}{273.15}\right)^{1.8}
\end{equation}
The drag coefficient depends on the Reynolds number
\begin{equation}
C_D = 24 (1 + 0.15 Re_p^{0.687})/Re_p
\end{equation}
which is given as 
\begin{equation}
Re_p = \frac{\rho_a v d}{\mu}
\end{equation}
where $\mu$ is the viscosity of air. \\
The particle acceleration is given by
\begin{equation}
\frac{\partial \vec{u_p}}{\partial t} = \frac{18 \mu C_D Re}{\rho_p d^2 24} \left(\vec{u} - \vec{u_p} \right) + \frac{\vec{g} (\rho_p - \rho)}{\rho_p}
\end{equation}

The first term on the right hand side is the drag force per mass. $\vec{u}$ is the velocity of the air, and $\vec{u_p}$ is the droplet velocity. $\mu$ is the viscosity of the fluid. A Discrete Random Walk model is used
to account for the effect of turbulent dispersion on the particle
trajectories. This includes the fluctuating component of the velocity
due to turbulence. 
The following velocity component is added to the local air velocity 
\begin{equation}
\vec{u}_{k} = \xi \sqrt{2k/3}
\end{equation}
where $\xi$ is a normally distributed random number that accounts for the randomness of turbulence about a mean value, and $k$ is the turbulent kinetic energy of the fluid \cite{redrow2011modeling, hathway2011cfd, li2018modelling}. 
The droplet positions are updated using the velocity verlet method, taking the fluid velocity from the linear interpolation of the lattice Boltzmann velocity field \cite{swope1982computer}.
\begin{figure}
	\begin{center}\includegraphics[width=0.5\linewidth]{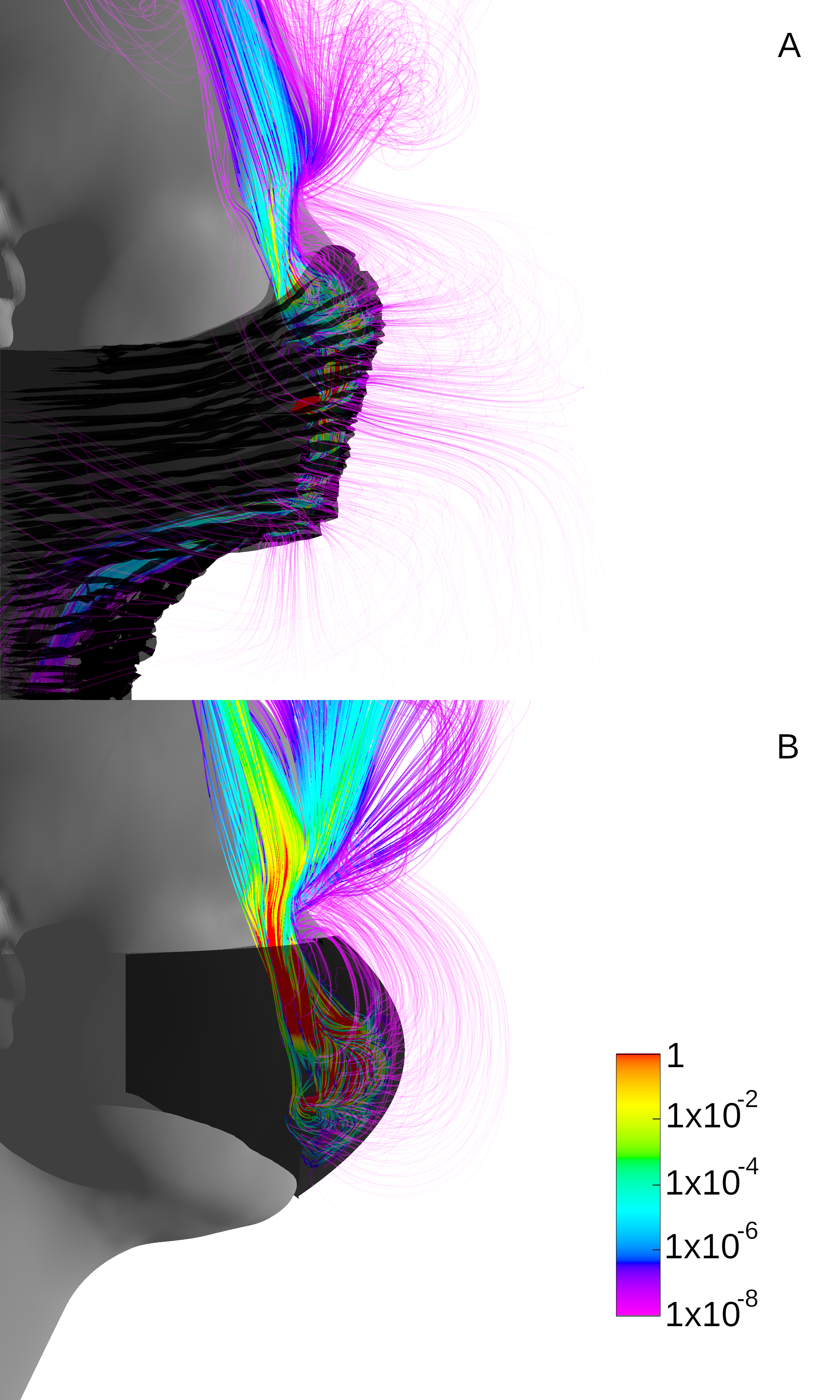}\end{center}
	\caption{Streamlines of fluid flow through a neck gaiter and a face mask. The velocity of the fluid, relative to the maximum velocity of 5 m/s, is depicted by the color of the streamlines.}
\end{figure}

\newpage
\section{Results}

We simulate the efficacy of a standard face mask, a neck gaiter and a face shield. The neck gaiter is captured using the Lattice Spring Model. The face mask geometry is designed to be a relatively good fit, with a small leakage on either side of the nose. The face shield is considered as an impermeable barrier that encircles the face.\\
The velocity at the mouth is set to 5 m/s and the neck gaiter has a porosity of 0.1 and a thickness 0.1 mm. 
The mask has a lower porosity and is thicker than the neck gaiter and the porosity is taken to be 0.05 and the thickness is 0.25 mm. 
The porosity of a node in the Lattice Boltzmann method is taken to be
\begin{equation}
p = \frac{L}{\sum (L_i/p_i)}
\end{equation}
where $L$ is the grid size, $L_i$ is the thickness of a layer (of either the mask or the air) and $p_i$ is the porosity of the layer (1 for just air). 
This assumes the layers of the mask are always perpendicular to the direction of flow, in terms of how the sub-grid porosity is handles, but allows us to calculate the flow of fluid through the permeable mask.\\

In Figure 1, the velocity of the particles are illustrated as they flow through and around the gaiter (Figure 1.A) and the mask (Figure 1.B). The face shield is not included as the velocity simply moves up and down along the face mask. The colors of the flow indicate the velocity of the particles with red being those particles moving the fastest and violet the slowest. 
The fluid simulations are ran for 25,000 time steps (a real time of 0.5 s) to establish typical a typical flow regime through and around the face coverings.
The neck gaiter allows some of the fluid to flow through the material, although the velocity is significantly reduced and spread out over the area of the neck gaiter rather than being focused at the mouth opening. In addition a large amount of fluid takes the path of least resistance and leaks through a gap between the nose and the cheek. As the neck gaiter wraps around the head, this is the only source of leakage.
The face mask, on the other hand, allows less of the fluid to permeate through the mask. The leakage is confined to a similar area to that of the neck gaiter, although in reality the areas of leakage might be expected to extend all the way around the mask, with leakage out of the sides of face masks common \cite{dbouk2020respiratory}.
Interestingly, the fluid flow around the face mask could cause infection to occur on the outside of the mask \cite{bae2020effectiveness}.\\

The emission of 10,000 respiratory droplets (with a uniform distribution between 1 $\mu$m and 100 $\mu$ m) at the same time is simulated given the flow profiles from the Lattice Boltzmann method. In Figure 2, the distribution of respiratory droplets after 0.2 s is illustrated for both the gaiter (Figure 1.A) and the mask (Figure 1.B). The larger droplets are represented in brighter colors (red and yellow) with smaller droplets illustrated in darker colors (violet). 
The filtration efficiency of the masks are assumed to be linear with a value of 0.5 for a radius of 0 and a filtration efficiency of 1 for droplets of radius 10 $\mu$m. The gaiters are considered to be less efficient at filtering the respiratory droplets. The filtration efficiency is 0 for a droplet radius of 0, and 1 for a droplet of radius 20 $\mu$m \cite{zangmeister2020filtration, aydin2020performance, bagherifiltration}.
Note that the larger droplets (reds and yellows) are less likely to be emitted as a person speaks, but they would also be expected to have a significantly large volume and carry more pathogens than small droplets.
Both fabrics are assumed to stop larger particles, but the mask has a higher filtration efficiency both in terms of stopping smaller droplets (whereas the neck gaiter is assumed to offer no filtration of very small droplets) and the fact that the mask stops all droplets larger than 10 $\mu$m (as opposed to the neck gaiter which only stops droplets larger than 20 $\mu$ m. This can be seen as the projection of smaller droplets through the neck gaiter. However, the mask redirects more the fluid flow through leakages between the mask and the wearers face. As smaller droplets are more likely to follow this fluid flow, there are a significant number of smaller droplets emitted around the face mask.\\
\begin{figure}
	\begin{center}\includegraphics[width=0.5\linewidth]{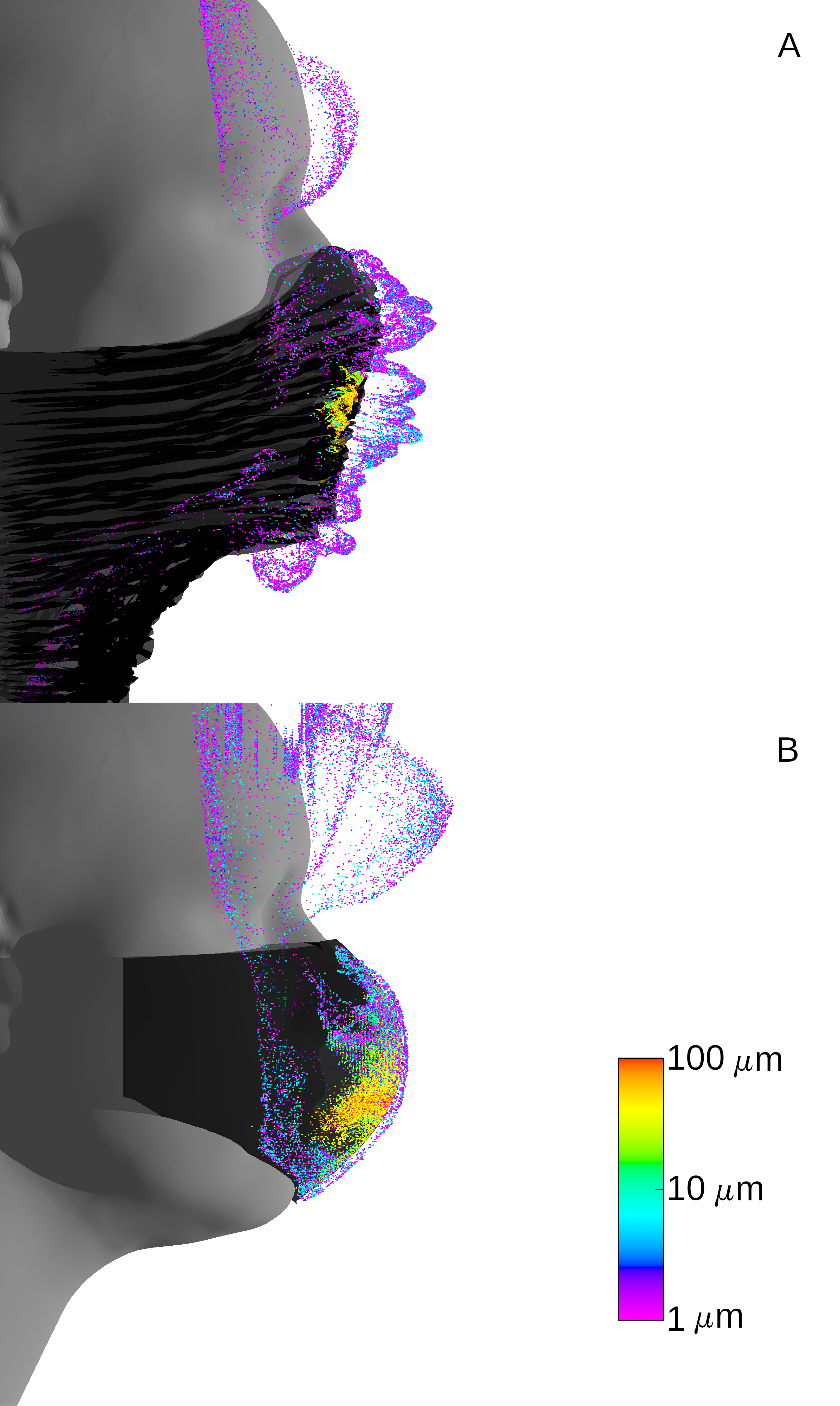}\end{center}
	\caption{The distribution of respiratory droplets after 0.2 s of being released from the mouth. The size of the droplets is depicted by the color of the droplets.}
\end{figure}

The fraction of droplets that leak around the face coverings is plotted as a function of droplet size in Figure 3a. 
This is the fraction of droplets that are neither collected nor pass through the face coverings, but rather move around the face coverings.
For comparison the face shield is included, that shows the smaller droplets move around the face shield and do not interact with the face shield at all. Larger droplets, that are much less likely to be emitted anyway, are less deflected by the surrounding air flow and impact the face shield (the leakage fraction goes to zero).
The mask is closer to the face than the face shield and does stop some droplets of all sizes, however for small droplets around 75\% of the respiratory droplets are emitted around the mask. This is because the mask is less permeable to air flow, and more air is directed through the leakage, taking the smaller droplets with this air flow.
The neck gaiter, however, is more permeable and an appreciable amount of flow passes through the neck gaier material. As a consequence, less than 20\% of the droplets (even for the smaller droplets) will leak from the mask without interacting with the neck gaiter in some form.\\

\begin{figure}
	\begin{center}\includegraphics[width=0.5\linewidth]{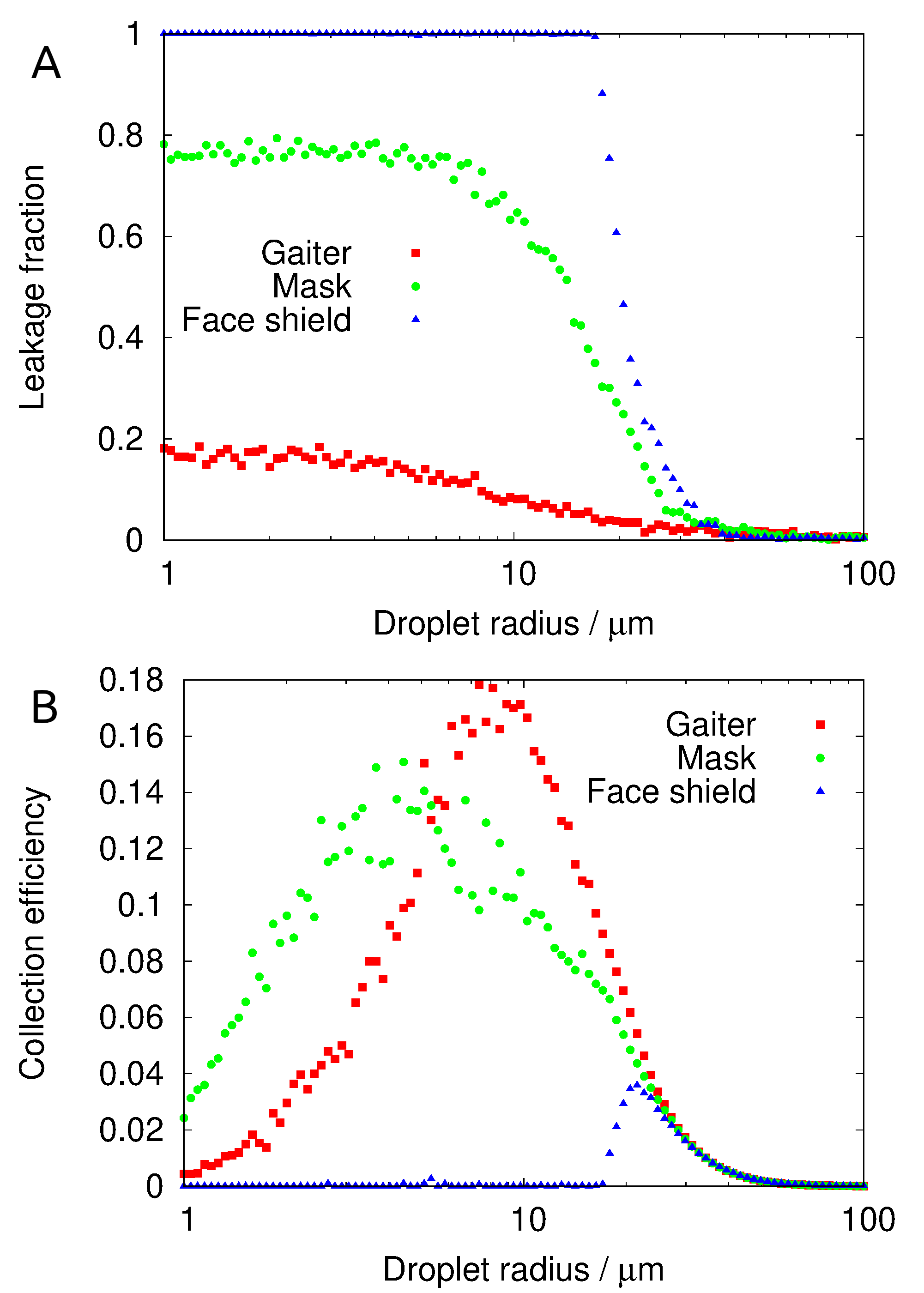}\end{center}
	\caption{a) The fraction of respiratory droplets that are transported around the mask, gaiter or face shield, as the air flow leaks out around the coverings, as a function of droplet size. b) The collection efficiency (fraction collected times probability of droplet expulsion) of the gaiter, mask and face shield as a function of respiratory droplet size.}
\end{figure}
The efficacy of the face masks are depicted in Figure 3c. In particular, the collection efficiency is plotted as a function of the droplet radius. Note that the collection efficiency is typically presented as the fraction of respiratory droplets collected by the face covering. Here, however, we present the fraction of respiratory droplets collected by the face covering relative to the distribution of droplets emitted. In other words, the fraction of droplets emitted at around 8 $\mu$m is taken to be 1 and the fraction at other sizes is taken from the distribution of Duguid \cite{duguid1946size}. Therefore, the collection efficiency goes to zero at very small or larger droplets as the probability of droplets being emitted at these sizes is much smaller. 
It can be seen that the collection efficiency for the face shield is very small, because all but the largest droplets follow the air flow around the face shield. 
The neck gaiter and the face mask interestingly have similar collection efficiencies.
For smaller droplets the face mask is more likely to allow these droplets to leak around the face mask, while the neck gaiter is more likely to allow these droplets to pass through the fabric.
The main difference is that at smaller droplets (around 1 $\mu$m) the neck gaiter offers very little filtration efficiency, while at medium sized droplets (around 10 $\mu$m) the mask will allow these to easily leak around the mask. 
Therefore, the neck gaiter might be expected to produce smaller droplets in a mist around the face covering, while the face mask will produce jets of air with slightly larger droplets directed out around the mask.\\

This is consistent with Drewnick \emph{et al.} who recently showed how leakage can have a dramatic effect on the efficacy of face coverings \cite{drewnick2020aerosol}. 
In addition, Lindsley \emph{et al.} found the same effects, and found that cloth face masks blocked 51\%, polyester neck gaiters blocked 47\%, and face shields blocked 2\% of respiratory droplets. 
Similar proportions are reported here, although it is worth noting that while the qualitative mechanism discussed are likely to be the same, the effects of fabric filtration efficiencies, porosity and the amount of leakage around the face coverings will play a large role.\\

\section{Summary and Conclusions}
At the outset of this manuscript, we presented a brief summary of the current state with respect to personal protective equipment and the mitigation attempts for the spread of COVID-19. The importance of facial coverings can not be overstated to mitigate the spread of COVID-19 in particular and airborne viruses in general. As mentioned, one study suggests that the economic benefit to society of each additional mask worn is upwards of \$6,000. However, in order to get broad compliance on the wearing of masks, they need to be economically accessible to all, comfortable, without increasing our impact on the environment. Cloth gaiters fit all of the aforementioned criteria - they are reusable, comfortable, and economically accessible.\\
	
We have demonstrated that neck gaiters may at least be as effective as more commonly accepted cotton face masks. In particular, we captured the fluid dynamics, and the transport of respiratory droplets, through and around various face coverings. We elucidated the role of droplet size on the collection efficiency and leakage fraction. We found that the neck gaiter was at least as effective as the cloth mask, and both were far superior to a face shield. The confirmation that the gaiter is as efficient as regular cloth masks may provides decision-makers with information to encourage individuals to use this intervention to help mitigate the spread of COVID-19. \\

\bibliographystyle{unsrt}
\bibliography{biblio}

\end{document}